\documentclass[letterpaper]{article}
\usepackage[T1]{fontenc}
\usepackage{geometry}
\usepackage{setspace}
\usepackage{achemso} 
\usepackage{graphicx}
\usepackage{float}
\newfloat{scheme}{htbp}{los}
\floatname{scheme}{Scheme}
\floatname{chart}{Chart}
\newfloat{graph}{htbp}{loh}
\usepackage{chemformula} 
\usepackage[version = 4]{mhchem} 
\usepackage{authblk}
\author[1,2]{Takahiro Uemura}
\author[1,3]{Yuto Moritake}
\author[2,4]{Eiichi Kuramochi}
\author[2,4]{Masaaki Ono}
\author[2,4]{Hisashi Sumikura}
\author[1,2,4]{Masaya Notomi*}
\affil[1]{Department of Physics, Institute of Science Tokyo, 2-12-1 Ookayama, Meguro-ku, Tokyo 152-8551, Japan}
\affil[2]{Basic Research Laboratories, NTT Inc, 3-1 Morinosato-Wakamiya, Atsugi-shi, Kanagawa 243-0198, Japan}
\affil[3]{Institute of Industrial Science, The University of Tokyo, 4-6-1 Komaba, Meguro-ku, Tokyo 153-8505, Japan}
\affil[4]{Nanophotonics Center, NTT Inc, 3-1, Morinosato-Wakamiya, Atsugi-shi, Kanagawa 243-0198, Japan}

\title{Low loss switchable topological photonic crystal enabled by submicron-scale patterning and phase-change of Sb\textsubscript{2}Se\textsubscript{3}}
\date{*Email: notomi@phys.sci.isct.ac.jp}

\begin{document}

\maketitle

\begin{abstract}
  Photonic topological insulators (PTIs) offer robust platforms for light manipulation, but reconfigurable control of their topological properties without degrading performance remains a major challenge. While phase-change materials (PCMs) provide large refractive index modulation, widely used materials such as Ge\textsubscript{2}Sb\textsubscript{2}Te\textsubscript{5} (GST) have been successfully deployed in commercial applications including optical data storage. However, they exhibit significant optical absorption in their crystalline state, which poses a challenge for transmissive photonic devices such as PTIs where high transparency is essential.
  Here, we overcome this fundamental limitation by integrating the ultra-low-loss PCM antimony triselenide (Sb\textsubscript{2}Se\textsubscript{3}) onto a silicon-based 2D PTI. We achieve submicron-scale selective patterning of Sb\textsubscript{2}Se\textsubscript{3} on a photonic crystal for the first time, and demonstrate a topological phase transition induced by the material phase change. Owing to the transparency of Sb\textsubscript{2}Se\textsubscript{3} in both its amorphous and crystalline states, a high $Q$-factor on the order of $10^3$ is preserved---representing nearly an order-of-magnitude improvement over previous GST-based devices. This work resolves the absorption-loss bottleneck in reconfigurable PTIs and paves the way for practical, low-loss, tunable topological photonic devices.
\end{abstract}

\section{Keywords}

Photonic topological insulators; Phase-change materials; Antimony triselenide; Photonic crystals; Nanofabrication

\section{Introduction}

Photonic topological insulators (PTIs), inspired by their counterparts in electronic systems \cite{xiao2010berry,hasan2010colloquium,qi2011topological,pesin2012spintronics}, have become a vibrant field of research due to their potential applications in robust optical signal processing \cite{lu2014topological}. In particular, two-dimensional (2D) PTIs, which support topologically protected edge modes that are immune to backscattering from structural disorders, are considered a key platform for next-generation photonic integrated circuits (PICs) \cite{wu2015scheme,barik2016two,barik2018topological,khanikaev2017two,hafezi2014topological,joannopoulos1997photonic,wang2009observation,susstrunk2015observation,haldane2008possible,fang2012realizing,khanikaev2013photonic,liang2013optical,rechtsman2013photonic}. However, a significant challenge remains: in conventional PTIs, the topological properties are determined by the as-fabricated nanostructure, typically a two-dimensional honeycomb-lattice dielectric photonic crystal based on crystalline C\textsubscript{6} symmetry \cite{wu2015scheme,barik2016two,barik2018topological,khanikaev2017two}. 
This makes it difficult to control the topological phase post-fabrication, and there are no established methods for actively tuning the topological properties.

Although various methods for tuning the refractive index, such as the thermo-optic effects of Silicon (Si) \cite{chong2004tuning}, the optical Kerr effect \cite{PhysRevB.70.205110}, carrier-induced nonlinearity \cite{Nozaki2010}, and photorefractive effects of chalcogenide glasses \cite{faraon2008local, eggleton2011chalcogenide}, have been explored, the achievable relative index change is typically small, on the order of $\Delta n / n \sim 10^{-3}$ to $10^{-2}$. In contrast, phase-change materials (PCMs) offer a significantly larger change in refractive index, for example a refractive-index contrast of $\Delta n/n \approx 0.65$ for Ge\textsubscript{2}Sb\textsubscript{2}Te\textsubscript{5} (GST) at 1.55\,$\mu$m ($n_\mathrm{a\text{-}GST}=4.39$, $\kappa_\mathrm{a\text{-}GST}=0.16$ for the amorphous phase and $n_{\mathrm{c\text{-}GST}}=7.25$, $\kappa_{\mathrm{c\text{-}GST}}=1.55$ for the crystalline phase \cite{tanaka2012ultra}), making them an attractive option for active control of photonic structures. PCMs are nonvolatile in both the amorphous and crystalline states (data retention exceeding 10 years) \cite{abdollahramezani2020tunable}. They can be switched either by thermal annealing at temperatures of about 150--170$^\circ$C \cite{nvemec2011optical} or by nanosecond-order optical or electrical pulses \cite{tanaka2012ultra, zheng2020nonvolatile}. Consequently, PCMs have been widely used in rewritable nanophotonic devices such as resonators \cite{rude2013optical, zheng2020nonvolatile}, modulators \cite{tanaka2012ultra, moriyama2014ultra, rios2015integrated}, optical switches and filters \cite{badri2021reconfigurable, xu2019low, morden2022tunable}, and reconfigurable metasurfaces \cite{chu2016active, nam2007electron, morden2022tunable}. 

Our group previously demonstrated the first experimental realization of a photonic topological phase transition by integrating GST nanostructures onto a 2D PTI \cite{doi:10.1126/sciadv.adp7779}. By leveraging the large refractive index contrast between the amorphous and crystalline states of GST, we successfully switched the topological invariant of the system. Although GST is a highly reliable material with established commercial manufacturing processes (e.g., in rewritable optical discs), its application to transmissive photonic devices is constrained by its significant optical absorption in the crystalline phase at telecommunication wavelengths. Ideally, PTIs are established on Hermitian systems composed of transparent dielectrics, where topological protection is strictly defined. Beyond this theoretical consideration, the material absorption caused an increase in the linewidth of the photonic band, rendering the crystalline phase unsuitable for low-loss, application-oriented devices.

In this work, we explore the potential of antimony selenide (Sb\textsubscript{2}Se\textsubscript{3}) to address this issue. While Sb\textsubscript{2}Se\textsubscript{3} is a relatively new material and its applications are still being developed, it has emerged as a highly promising material for reconfigurable photonics thanks to its unique combination of substantial refractive-index contrast ($\Delta n \approx 0.7$--$0.8$, $n_\mathrm{a\text{-}Sb_2Se_3}=3.29$ for the amorphous phase and $n_{\mathrm{c\text{-}Sb_2Se_3}}=4.05$ for the crystalline phase at 1.55\,$\mu$m) and, most importantly, ultra-low optical loss ($\kappa < 10^{-5}$) in both states at near-infrared wavelengths \cite{https://doi.org/10.1002/adfm.202002447, https://doi.org/10.1002/apxr.202400080}. The phase-transition temperature of Sb\textsubscript{2}Se\textsubscript{3} is around 190$^\circ$C, and experimental demonstrations of localized phase transitions using lasers have been reported \cite{Lawson:24}. These superior properties have enabled demonstrations of various high-performance nonvolatile photonic components, such as programmable optical switches \cite{doi:10.1126/sciadv.abg3500, 10.1063/5.0234637}, metasurfaces \cite{doi:10.1021/acs.nanolett.4c04904}, microring resonators \cite{doi:10.1021/acs.nanolett.3c03353}, and even platforms for free-form, rewritable photonic circuits using laser-based techniques \cite{doi:10.1126/sciadv.adk1361}. Since Sb\textsubscript{2}Se\textsubscript{3} maintains its ultra-low-loss property in both states, it is expected to enable effective band modulation when integrated with optical nanostructures such as photonic crystals. However, to date, there have been no reports demonstrating this capability. In this context, a critical challenge is the precise nanoscale patterning and positioning of Sb\textsubscript{2}Se\textsubscript{3} on photonic structures. While submicron-scale patterning of Sb\textsubscript{2}Se\textsubscript{3} has been demonstrated in Ref.\cite{doi:10.1021/acs.nanolett.3c03353}, integrating such patterns onto photonic crystals with the required precision remains significantly challenging.

This study aims to establish the feasibility of Sb\textsubscript{2}Se\textsubscript{3} nanopatterning at photonic crystal dimensions and to demonstrate absorption-free topological phase transitions that preserve high $Q$-factors throughout the switching process. The key to our approach is selective tuning of specific eigenmodes through spatially localized patterning---by placing the phase-change material where targeted modes have high field intensity while others do not, we achieve the necessary symmetry-selective frequency shift for topological inversion. This makes submicron-scale patterning of Sb\textsubscript{2}Se\textsubscript{3} essential to our scheme. We present the design principles for integrating Sb\textsubscript{2}Se\textsubscript{3} nanopatterns with PTI structures, followed by the development of nanofabrication processes that achieve reliable phase-change operation at submicron scales---representing the first successful demonstration of such capability with this emerging material onto photonic crystals. We provide clear evidence of topological band inversion with dramatically enhanced spectral clarity compared to absorptive PCMs. Quantitative analysis reveals a nearly order-of-magnitude improvement in crystalline-state $Q$-factors compared to GST-based devices, effectively resolving the fundamental absorption limitation.
Beyond establishing Sb\textsubscript{2}Se\textsubscript{3} as a viable platform for topological applications, this work opens new possibilities for nanophotonic devices requiring precise refractive index control without absorption penalties. The demonstrated nanoscale patterning and phase-change control capabilities extend well beyond topological systems, offering new opportunities for advanced reconfigurable optical circuits. 

\section{Design of high-Q reconfigurable topological photonic crystals}

\begin{figure}[htbp]
	\centering
	\includegraphics[width=1.0\linewidth]{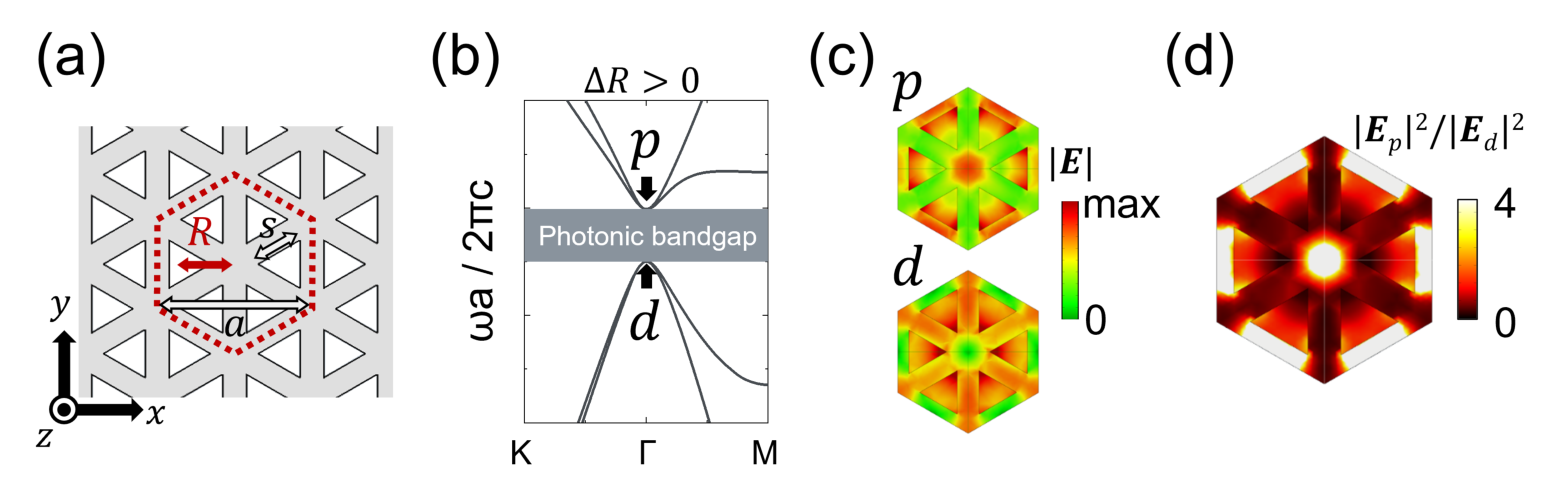}
	\caption{
		(a) Schematic of a Barik-type Si topological photonic crystal slab. 
        (b) Band dispersion curves of the TE-like mode for the photonic crystal with $\Delta R > 0$.
		(c) Electric-field distributions of the $p$- and $d$-modes at the $\Gamma$ point in the unit cell. The plotted field distributions represent the sum of two eigenmodes of the effective Hamiltonian at the $\Gamma$ point. 
        (d) ${\lvert \boldsymbol{E}_p \rvert }^2 / {\lvert \boldsymbol{E}_d \rvert }^2$ distribution in the unit cell.
		}
	\label{fig:fig1}
\end{figure}

Here, we present the design of our topological photonic crystal slab and its fundamental optical properties. All the numerical simulations in this study were performed using a three-dimensional finite-element method (FEM, COMSOL Multiphysics). 

Our topological photonic crystal slab is based on the Barik-type design \cite{barik2016two,barik2018topological}, which consists of a silicon slab with triangular holes arranged in a honeycomb lattice, as shown in Fig.~\ref{fig:fig1}(a). See also Ref.~\cite{doi:10.1126/sciadv.adp7779} for details; the following is a brief explanation. We start with a unit cell with six holes, corresponding to a tripled primitive unit cell of the honeycomb lattice. The hole-position shift $\Delta R$ is defined as $\Delta R = R - a/3$, where $R$ is the distance from the center of the unit cell to the center of a triangle, and $a$ is the lattice constant. By varying $\Delta R$, we can induce a topological phase transition between trivial ($\Delta R < 0$) and non-trivial ($\Delta R > 0$) phases. 
The photonic crystal slab supports transverse-electric (TE)--like modes, where the electric field is primarily polarized in the plane of the slab. The band dispersion curves of the TE-like mode with $\Delta R > 0$ are shown in Fig.~\ref{fig:fig1}(b). The band structure exhibits a complete photonic bandgap between the $p$- and $d$-modes. The photonic topological properties of this system are determined by the frequencies of the $p$- and $d$-modes at the $\Gamma$ point. When $\Delta R < 0$, the $p$-mode has a lower frequency than the $d$-mode, indicating a trivial phase. Conversely, when $\Delta R > 0$, the $d$-mode has a lower frequency than the $p$-mode, indicating a non-trivial phase. This band inversion between the $p$- and $d$-modes leads to a topological phase transition.

The key concept of our study is to utilize the phase-change material Sb\textsubscript{2}Se\textsubscript{3} to induce a topological phase transition by changing the effective hole shift through the refractive-index contrast of a- and c-Sb\textsubscript{2}Se\textsubscript{3}. The electric-field distributions of the $p$- and $d$-modes at the $\Gamma$ point are shown in Fig.~\ref{fig:fig1}(c), where $\boldsymbol{E}_p := \boldsymbol{E}_{p_x} + \boldsymbol{E}_{p_y}$ and $\boldsymbol{E}_d := \boldsymbol{E}_{d_{xy}} + \boldsymbol{E}_{d_{x^2-y^2}}$. Here, $\boldsymbol{E}_{p_x}$, $\boldsymbol{E}_{p_y}$, $\boldsymbol{E}_{d_{xy}}$, and $\boldsymbol{E}_{d_{x^2-y^2}}$ represent the electric-field distributions of the bases of the eigenmodes at the $\Gamma$ point. Figure~\ref{fig:fig1}(d) shows the spatial distribution of ${\lvert \boldsymbol{E}_p \rvert }^2 / {\lvert \boldsymbol{E}_d \rvert }^2$ in the unit cell. The figure indicates that the $p$-mode is mainly localized in the region between the two triangles, while the $d$-mode is localized around the center of the unit cell. This difference in localization leads to a band inversion when transitioning from $\Delta R > 0$ to $\Delta R < 0$, resulting in a topological phase transition.

\begin{figure}[htbp]
	\centering
	\includegraphics[width=0.7\linewidth]{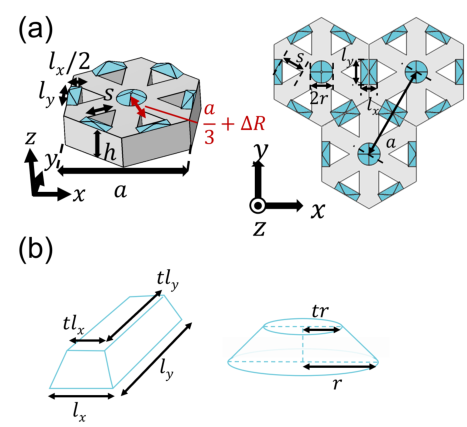}
	\caption{
		(a) The unit cell of the designed Sb\textsubscript{2}Se\textsubscript{3}-loaded PhC structure for numerical simulation. The left side shows the unit cell with a tapered rectangular Sb\textsubscript{2}Se\textsubscript{3} pattern, and the right side defines the structural variables. (b) The geometry of tapered rectangular and cylindrical Sb\textsubscript{2}Se\textsubscript{3} patterns with reduced upper parts.
		}
	\label{fig:fig2}
\end{figure}

Figure~\ref{fig:fig2}(a) illustrates the geometry of the Sb\textsubscript{2}Se\textsubscript{3}-loaded PTI used for the numerical simulations.
In the subsequent numerical simulations and experiments, the lattice constant $a$ and the slab thickness $h$ were set to 740\,nm and 205\,nm, respectively, while the side length of a triangle $s$ was chosen as $0.24a$. The silicon slab has a refractive index of $n_{\text{Si}} = 3.48$, and it is surrounded by air ($n_{\text{air}} = 1$) on the top and a SiO\textsubscript{2} substrate ($n_{\text{SiO}_2} = 1.45$) on the bottom. 
To selectively perturb the $p$-mode frequency, 50-nm-thick Sb\textsubscript{2}Se\textsubscript{3} nanopatterns are strategically placed in regions where the electric-field intensity contrast between the $p$- and $d$-modes is maximized as seen in Fig.~\ref{fig:fig1}(d). The key geometric variables for the simulation are labeled, including the lattice constant $a$, the triangle side length $s$, and the dimensions of the circular ($r=(4/45)a$) and rectangular ($l_x=a - 2(R+s\sqrt{3})$, $l_y=s$) Sb\textsubscript{2}Se\textsubscript{3} patterns. The topological properties are tuned by adjusting the hole position, which is defined by the shift parameter $\Delta R$. The principle of our design is to balance the effects of this geometric hole shift with an effective refractive-index modulation from the periodically patterned phase-change material. This material patterning effectively acts as a negative hole shift ($\Delta R < 0$), the magnitude of which depends on the material phase. Therefore, by introducing a positive geometric shift ($\Delta R > 0$) in the fabricated structure, we can precisely balance these two competing effects. This configuration allows the material phase transition to tip the balance, invert the frequency ordering of the $p$- and $d$-modes, and thereby achieve a topological phase transition.
To accurately replicate the fabricated device, the simulations incorporate a tapered profile for the Sb\textsubscript{2}Se\textsubscript{3} nanostructures, as depicted in Fig.~\ref{fig:fig2}(b). This tapered geometry, which accounts for the sloped sidewalls of the patterns, was determined based on atomic-force microscope (AFM) measurements of the fabricated structures, as shown in Fig.~\ref{fig:fig4}(c) in the next section.

\begin{figure}[htbp]
	\centering
	\includegraphics[width=1.0\linewidth]{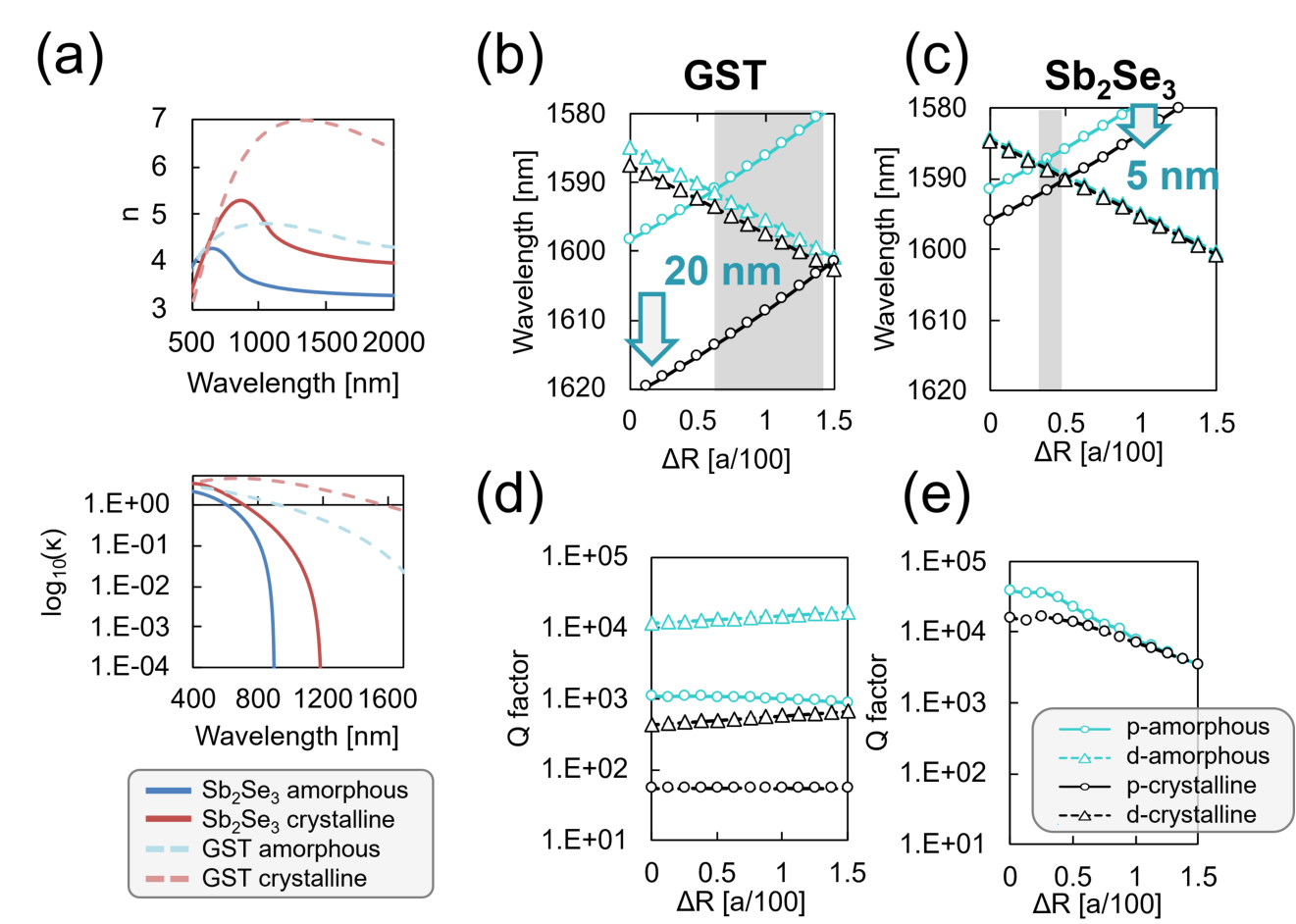}
	\caption{
		(a) Refractive index and extinction coefficient of the 80-nm Sb\textsubscript{2}Se\textsubscript{3} and 30-nm GST films obtained from ellipsometry.
		(b), (c) Mode frequencies of the $p$- and $d$-modes at the $\Gamma$ point as a function of hole shift $\Delta R$ for (b) GST-loaded PhC and (c) Sb\textsubscript{2}Se\textsubscript{3}-loaded PhC. Solid and broken lines represent the $p$- and $d$-modes, respectively. Blue and black lines indicate the amorphous and crystalline states, respectively. 		(d), (e) Quality factors of the $p$- and $d$-modes at the $\Gamma$ point as a function of $\Delta R$ for (d) GST-loaded PhC and (e) Sb\textsubscript{2}Se\textsubscript{3}-loaded PhC. The $d$-mode Q-factors are not shown in (e) because they are theoretically infinite in the absence of material absorption.
		}
	\label{fig:fig3}
\end{figure}

The refractive index $n$ and extinction coefficient $\kappa$ are plotted as a function of wavelength for an 80-nm-thick Sb\textsubscript{2}Se\textsubscript{3} film and a 30-nm-thick GST film (reference), both deposited on Si substrates via RF sputtering. These optical constants were determined by fitting spectroscopic ellipsometry data using the Cody--Lorentz model. The optical constants obtained here were used as material parameters in the numerical simulations for this study. At the telecommunication wavelength of $\lambda = 1{,}550$\,nm, the complex refractive index for GST was measured to be $n_{\mathrm{a\text{-}GST}}=4.33$ and $\kappa_{\mathrm{a\text{-}GST}} = 0.07$ for the amorphous state, and $n_{\mathrm{c\text{-}GST}} = 6.93$ and $\kappa_{\mathrm{c\text{-}GST}} = 1.02$ for the crystalline state. In contrast, Sb\textsubscript{2}Se\textsubscript{3} was found to be $n_{\mathrm{a\text{-}Sb_2Se_3}}=3.34$ (amorphous) and $n_{\mathrm{c\text{-}Sb_2Se_3}}=4.32$ (crystalline). Notably, the extinction coefficient of Sb\textsubscript{2}Se\textsubscript{3} is negligible in both states at this wavelength, maintaining a low-loss characteristic ($\kappa < 10^{-4}$). The measured refractive index and extinction coefficient are in good agreement with previous studies \cite{https://doi.org/10.1002/adfm.202002447}. This refractive-index contrast allows effective tuning of the photonic band structure without introducing material absorption.

Figures~\ref{fig:fig3}(b) and \ref{fig:fig3}(c) show the simulated band wavelengths of the $p$-mode (circles) and $d$-mode (triangles) at the $\Gamma$ point for both GST- and Sb\textsubscript{2}Se\textsubscript{3}-loaded PTIs shown in Fig.~\ref{fig:fig2}(a), as a function of the hole-shift parameter $\Delta R$. The sky-blue and black lines represent the modes in the amorphous and crystalline states, respectively. The parameter $\Delta R$ sets the initial frequency separation between these modes. The material phase transition from the amorphous (sky-blue lines) to the crystalline (black lines) state induces a selective redshift in the $p$-mode, attributed to the increase in the material refractive index. For GST, this shift is substantial, approximately 20\,nm. For Sb\textsubscript{2}Se\textsubscript{3}, the shift is smaller, around 5\,nm, due to its lower refractive-index contrast ($\Delta n$). The gray-shaded region in each plot highlights where the redshift of the $p$-mode is sufficient to invert the ordering of the $p$- and $d$-mode frequencies, implying that a topological phase transition can be induced by the material phase transition. The simulations therefore confirm that, by choosing an appropriate initial $\Delta R$, a topological phase transition can be induced in both material systems. 

The key advantage of Sb\textsubscript{2}Se\textsubscript{3} is evident in the high Q-factor. For the GST-loaded PTI, the Q-factor of the $p$-mode suffers severe degradation, dropping by more than an order of magnitude from over $10^3$ to below $10^2$ upon crystallization [Fig.~\ref{fig:fig3}(d)], directly resulting from high material absorption in crystalline GST. In contrast, the $p$-mode of the
Sb\textsubscript{2}Se\textsubscript{3}-loaded PTI maintains a high Q-factor on the order of $10^4$ in both amorphous and crystalline states [Fig.~\ref{fig:fig3}(e)]. Notably, the Q-factor data for the $d$-mode are not shown in Fig.~\ref{fig:fig3}(e), because the $d$-mode, owing to its quadrupolar symmetry, does not radiate to the far field in the absence of material absorption. In an ideal lossless system, the $d$-mode would be completely dark at the $\Gamma$ point, and its Q-factor would be theoretically infinite. This clearly demonstrates that the ultra-low-loss nature of Sb\textsubscript{2}Se\textsubscript{3} allows control of the topological phase without significant performance penalties, making it a strong candidate for reconfigurable PTIs.

\section{Fabrication}

\begin{figure}[htbp]
	\centering
	\includegraphics[width=1.0\linewidth]{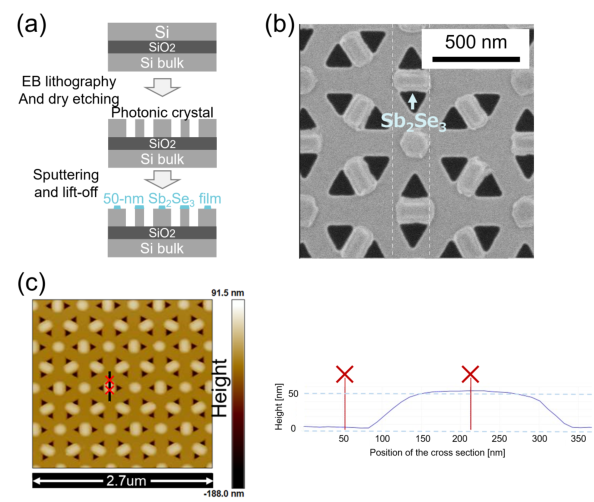}
	\caption{
		(a) Fabrication process of the Sb\textsubscript{2}Se\textsubscript{3}-loaded PTI. (b) The designed Sb\textsubscript{2}Se\textsubscript{3}-loaded PhC structure and a SEM image of the fabricated PhC. The lattice constant and slab thickness are $a = 740$\,nm and $h = 205$\,nm, respectively.
		(c) AFM image and cross-sectional profile of the deposited Sb\textsubscript{2}Se\textsubscript{3} nanostructure. The height of the Sb\textsubscript{2}Se\textsubscript{3} structure is approximately 50\,nm.
		}
	\label{fig:fig4}
\end{figure}

The designed PTIs were fabricated on a silicon-on-insulator (SOI) wafer with a 205-nm-thick top Si layer. The fabrication process is schematically illustrated in Fig.~\ref{fig:fig4}(a). First, the photonic crystal (PhC) patterns were defined in a resist layer using electron-beam (EB) lithography. The patterns were subsequently transferred into the top Si layer through dry etching to form the PhC slab. Following the PhC fabrication, a second EB-lithography step was performed to define the regions for the Sb\textsubscript{2}Se\textsubscript{3} nanostructures. An Sb\textsubscript{2}Se\textsubscript{3} film with a thickness of approximately 50\,nm was then deposited by RF sputtering using an alloy target. The deposition was carried out in an Ar atmosphere at a pressure of $3.0 \times 10^{-1}$\,Pa, with a deposition rate of 0.15\,nm/s. The patterned Sb\textsubscript{2}Se\textsubscript{3} nanostructures were fabricated via a lift-off process. Figure~\ref{fig:fig4}(b) presents a scanning electron microscope (SEM) image of a fabricated device, confirming precise alignment of the Sb\textsubscript{2}Se\textsubscript{3} nanostructures with the underlying Si PhC. The topography and thickness of the deposited film were further characterized using an atomic-force microscope (AFM), as shown in Fig.~\ref{fig:fig4}(c). The cross-sectional profile confirms a film thickness of approximately 50\,nm. The deposited Sb\textsubscript{2}Se\textsubscript{3} structures are not perfectly rectangular; instead, the upper part along the $z$-direction is rounded into a dome shape. This morphology has no significant effect on device performance but is considered in the numerical simulations, as illustrated in Fig.~\ref{fig:fig2}(b). After initial optical measurements of the as-deposited (amorphous) samples, the devices were crystallized by thermal annealing on a hot plate at 200$^\circ$C for 10 minutes.

\section{Experimental results}

\subsection{Experimental setup}

\begin{figure}[htbp]
	\centering
	\includegraphics[width=0.7\linewidth]{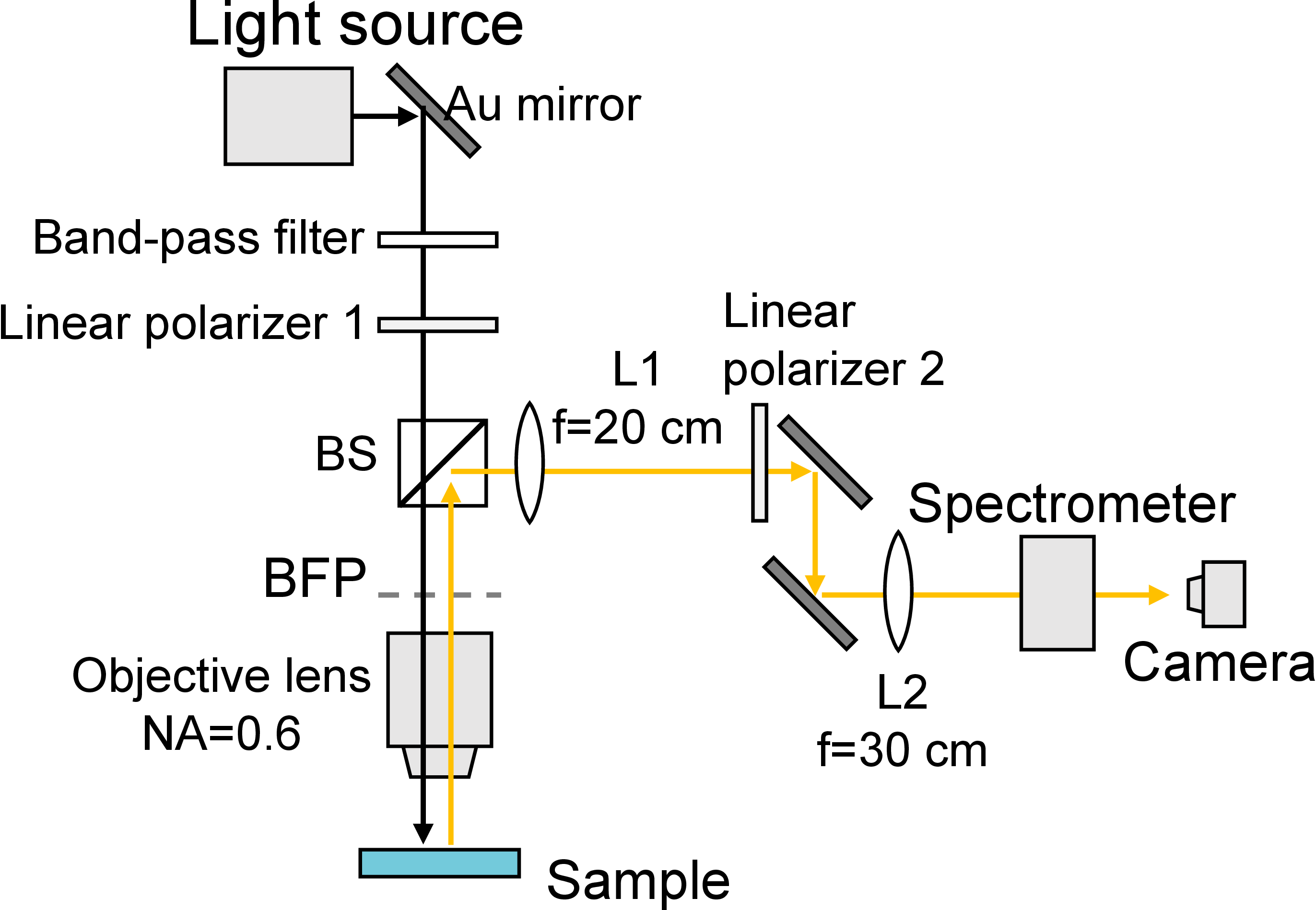}
	\caption{
		Experimental setup for the reflection measurement.
		}
	\label{fig:fig5}
\end{figure}

Figure~\ref{fig:fig5} illustrates the experimental arrangement used for the band measurements. Broadband, unpolarized light from a tungsten--halogen lamp (Thorlabs, SLS302) was focused onto the sample with an aspherical lens (Thorlabs, C105TMD-C, NA = 0.6). The lens back focal plane was relayed onto the entrance slit of an imaging spectrometer (Teledyne Princeton Instruments, IsoPlane 320) equipped with an InGaAs camera (NIRvana HS) by a pair of convex lenses (L1, L2) with focal lengths of 20 and 30\,cm, respectively. The polarization state of the incident light was defined by linear polarizer~1, with linear polarizer~2 oriented orthogonally, which allows us to detect only the resonantly scattered light from the photonic crystal slab. Narrowband bandpass filters (Thorlabs FBH series, FWHM = 12\,nm) were inserted to isolate specific frequencies. 

\subsection{Band measurement}

\begin{figure}[htbp]
	\centering
	\includegraphics[width=1 \linewidth]{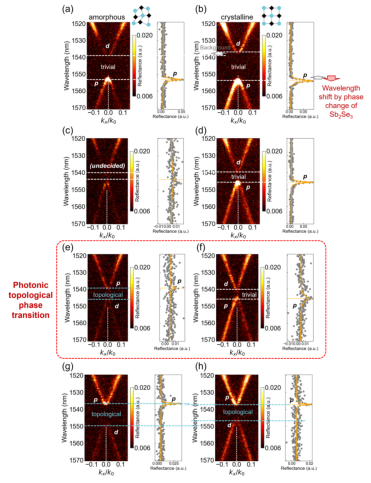}
	\caption{
		Measured reflection spectra along the $k_x$ direction and the cross section of the measured band diagram at the $\Gamma$ point. (a), (b) $\Delta R = 0$ (trivial phase). (c), (d) $\Delta R = 0.005a$ (from indeterminate to trivial phase). (e), (f) $\Delta R = 0.00675a$ (phase transition). (g), (h) $\Delta R = 0.00875a$ (topological phase). (a), (c), (e), (g) correspond to the amorphous phase, and (b), (d), (f), (h) correspond to the crystalline phase. The orange line on the cross-section plots represents the fitting curve based on temporal coupled-mode theory.
		}
	\label{fig:fig6}
\end{figure}

Figure~\ref{fig:fig6} shows the measured band dispersion relations for samples with different hole shifts ($\Delta R$) before and after thermal annealing, corresponding to the amorphous and crystalline states of Sb\textsubscript{2}Se\textsubscript{3}, respectively. To the right of each band diagram, we present the cross-section of the reflection spectrum at the $\Gamma$ point ($k_x=0$) along with its fitting curve, described in Eq.~(\ref{def_TCMT}) below. The wavenumber integration width for the cross-section is set to $\Delta k_x/k_0 = 0.0012$. The fundamental principle for interpreting the measured spectra lies in the symmetry of the photonic-crystal modes. The $p$- and $d$-modes possess different symmetries; specifically, the $d$-mode has a quadrupolar symmetry, which makes it unobservable at the $\Gamma$ point. In other words, due to the mismatch between the mode symmetry and that of free space, it does not radiate to the far field and thus cannot be observed. Consequently, the bright band observed at the $\Gamma$ point can be identified as the $p$-mode. A topological (non-trivial) phase is defined by the condition $\omega_p > \omega_d$, while a trivial phase is defined by $\omega_p < \omega_d$.

First, the sample with $\Delta R = 0$ [Figs.~\ref{fig:fig6}(a),(b)] serves as a control, designed to remain in the trivial phase. In both the amorphous (a) and crystalline (b) states, the lower band is bright at the $\Gamma$ point, confirming the trivial state ($\omega_p < \omega_d$). Importantly, the spectral lines are very sharp in both states, indicating that a high Q-factor is maintained. Here, the thermal annealing process causes slight surface oxidation of the silicon substrate, leading to a uniform blueshift of the entire spectrum. The size of blueshift is determined to be 2.3\,nm by measuring a reference sample without Sb\textsubscript{2}Se\textsubscript{3} loading and observing the frequency shift of the $p$-mode at the $\Gamma$ point. 
Next, for $\Delta R = 0.005a$ [Figs.~\ref{fig:fig6}(c),(d)], we observe a state just before the phase transition. In the amorphous phase (c), both bands are very faint near the $\Gamma$ point. This suggests that the system is approaching an unperturbed state with enhanced symmetry, which suppresses out-of-plane radiation loss. Upon transitioning to the crystalline phase (d), the lower band becomes clearly visible, indicating a trivial state ($\omega_p < \omega_d$). We conclude from these results that this sample is on the cusp of a topological transition in its amorphous phase.
The central result demonstrating the phase transition is from the sample with $\Delta R = 0.00675a$ [Figs.~\ref{fig:fig6}(e),(f)]. In the amorphous phase (e), the upper band is bright at the $\Gamma$ point while the signal from the lower band vanishes. This clearly indicates a topological phase with $\omega_p > \omega_d$. In contrast, after annealing, the situation is inverted in the crystalline phase (f): the lower band becomes bright, and the upper band disappears. This signifies a transition to the trivial phase ($\omega_p < \omega_d$), providing the first experimental demonstration of a photonic topological phase transition induced by the phase change of Sb\textsubscript{2}Se\textsubscript{3}.
Finally, the sample with $\Delta R = 0.00875a$ [Figs.~\ref{fig:fig6}(g),(h)] is a control designed to remain in the topological phase. As expected, the upper band is bright in both the amorphous (g) and crystalline (h) states, maintaining the topological condition ($\omega_p > \omega_d$).

Importantly, the band diagrams in Fig.~\ref{fig:fig6} reveal sharp, well-defined spectral features in the crystalline state that would be impossible to achieve with GST-based systems. In particular, the crystalline-state spectra [panels (b), (d), (f), (h)] exhibit narrow linewidths and high contrast that enable unambiguous identification of topological phase boundaries---a capability that was compromised in previous absorptive systems.
This enhanced spectral clarity is not merely a qualitative leap but represents a quantitative improvement in our ability to characterize and control topological phases. The cross-section data at the $\Gamma$ point [Fig.~\ref{fig:fig6}(e),(f)] demonstrate this advancement most clearly: the transition from a bright upper band to a bright lower band provides unambiguous evidence of band inversion, with spectral features that remain sharp and well-resolved throughout the phase transition. 

\begin{figure}[htbp]
	\centering
	\includegraphics[width=0.9\linewidth]{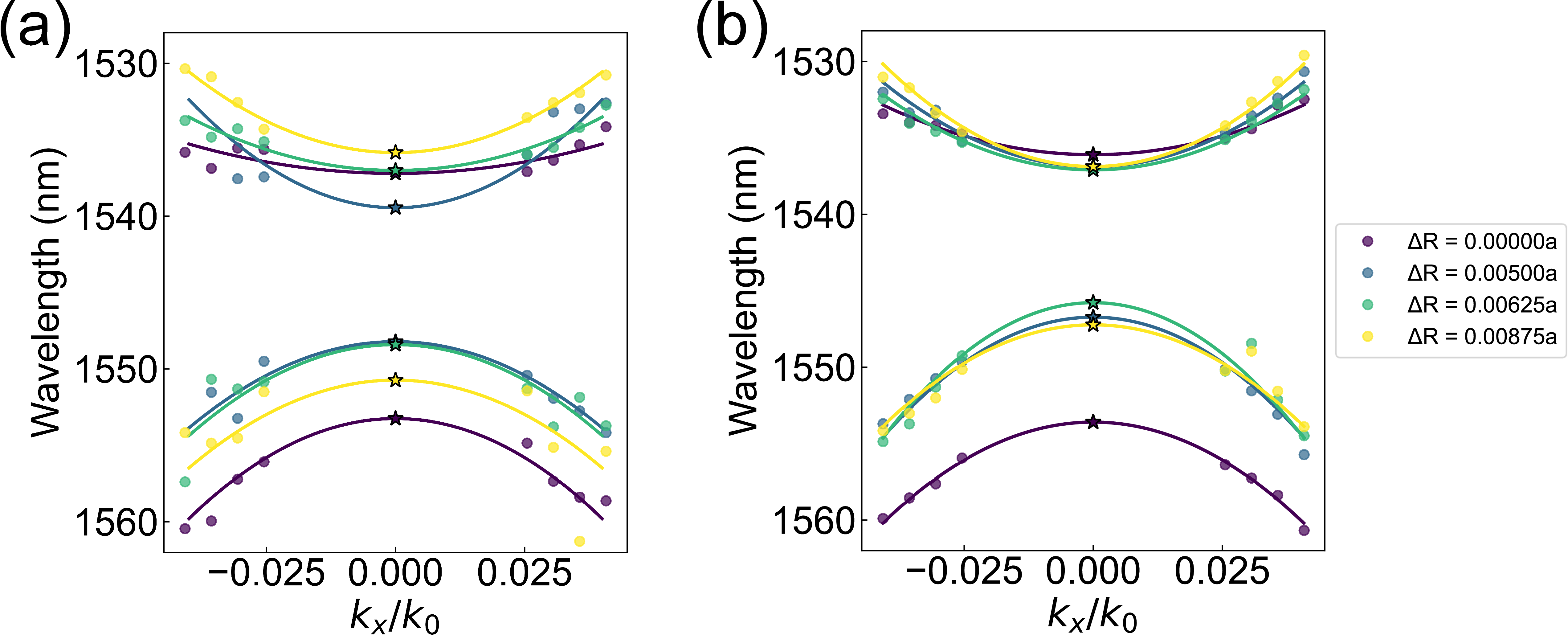}
	\caption{
		Resonant wavelengths at $|k_x|/k_0 \in [0.025, 0.04]$ and their parabolic fitting curves for $\Delta R = 0, 0.005a, 0.00675a, 0.00875a$ in (a) amorphous and (b) crystalline states.
		}
	\label{fig:fig7}
\end{figure}

To quantitatively assess the primary advantage of using Sb\textsubscript{2}Se\textsubscript{3}, we analyzed the Q-factor of the $p$-mode at the $\Gamma$ point by fitting the measured reflection spectra with temporal coupled-mode theory (TCMT) \cite{Hsu2013}. 
\begin{align}
    \label{def_TCMT}
    R(\omega) = {\left\lvert r'(\omega) + \frac{Q_{u}^{-1} \, e^{i2\theta}}{2i\!\left(1-\frac{\omega}{\omega_p}\right) - Q_{\mathrm{tot}}^{-1}} \right\rvert }^2 + a\omega + b
\end{align}
Here, the resonant frequency is denoted by $\omega_p$, while the total attenuation coefficient is $\gamma=\omega_p / (2Q_{\mathrm{tot}})$. The coupling coefficient is $d=\sqrt{2\gamma_{u}}\,e^{i\theta}$, and the attenuation coefficient by upper out-of-plane radiation is $\gamma_{u}=\omega_p/(2Q_{u})$. 
$Q_{u}$ and $Q_{\mathrm{tot}}$ represent the upper out-of-plane radiation Q and the total Q, respectively.
$Q_{u}$ in Eq.~\eqref{def_TCMT} approximately corresponds to the amplitude (brightness) of the spectrum, while $Q_{\mathrm{tot}}$ corresponds to the linewidth.
Moreover, although $r'(\omega)$ accounts for non-resonant reflection by a dielectric multilayer, we set $r'(\omega)=0$ in the fitting because the two linear polarizers are orthogonal in the experiment and only the resonant component is observed. Therefore, the fitting parameters were $\omega_p, Q_{\mathrm{tot}}, Q_{u}$, and $\theta$.

In practice, fitting the reflection spectrum at $k_x=0$ faces difficulties by the dark nature of the $d$-mode, which does not radiate at the $\Gamma$ point in an ideal lossless system. To determine the $\Gamma$-point wavelengths of both bands, we therefore extract the resonance peaks at small finite in-plane wavevectors and extrapolate to $k_x=0$. Specifically, we fit the peaks measured over $0.025 \le |k_x|/k_0 \le 0.04$ with a sampling step of $\Delta(|k_x|/k_0)=0.005$. Since the band dispersion is even in $k_x$, it can be approximated near $\Gamma$ by a parabola,
\begin{align}
    \label{def_parabolic_fit}
	f(k_x) = \lambda_0 + \lambda_2 k_x^2
\end{align}
and we use the fitted $\lambda_0$ as the $\Gamma$-point wavelength. Figures~\ref{fig:fig7}(a) and \ref{fig:fig7}(b) plot the measured resonant wavelengths at finite $|k_x|$ together with their parabolic fits for each $\Delta R$ in the amorphous and crystalline states, respectively. Data points where the Lorentzian fitting failed due to weak spectral intensity were excluded from the analysis; In the amorphous state of Fig.~\ref{fig:fig7}(a), the slices at $k_x/k_0=-0.04$ and $-0.035$ could not be fitted reliably and are therefore not shown. The crystalline state generally exhibits smaller $Q_u$ and thus stronger signals than the amorphous state, resulting in more precise peak extraction in the data points. The extrapolated $\Gamma$-point wavelengths $\lambda_0$ are indicated by star symbols in the plots, demonstrating that the parabolic fitting captures the band dispersion near the $\Gamma$ point.

\begin{figure}[htbp]
	\centering
	\includegraphics[width=1\linewidth]{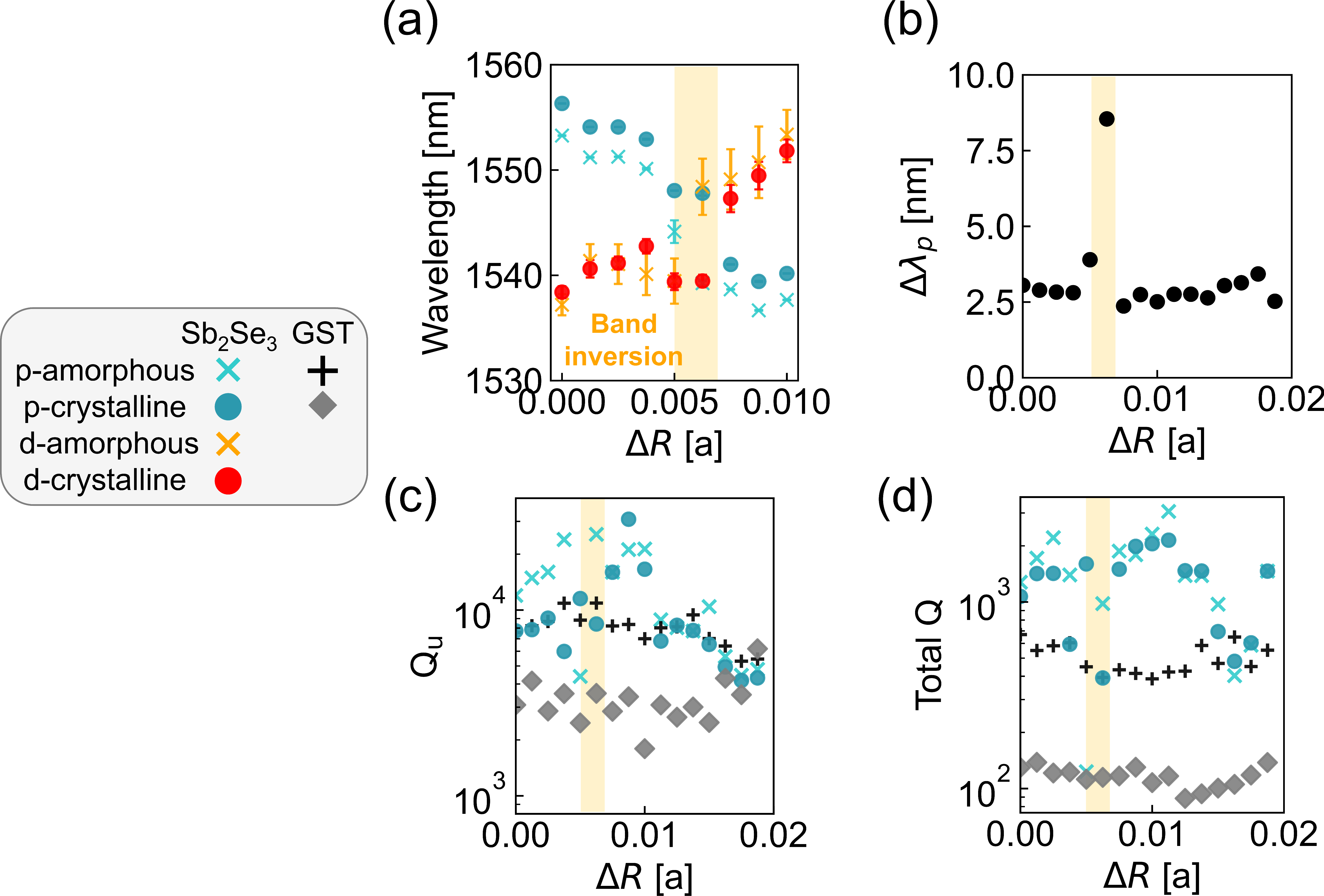}
	\caption{
		(a) Resonant wavelengths at the $\Gamma$ point, (b) wavelength shift of the $p$-mode through the material phase transition, (c) out-of-plane radiation Q-factor ($Q_u$), and (d) total Q-factor ($Q_\mathrm{total}$) as a function of the hole-shift parameter $\Delta R$. 
		The black and gray dots in (c) and (d) represent the data for GST as a reference \cite{doi:10.1126/sciadv.adp7779}.
		The orange-shaded region in (a) indicates the range where a topological phase transition can be induced by the phase change of Sb\textsubscript{2}Se\textsubscript{3}.
		}
	\label{fig:fig8}
\end{figure}

Figure~\ref{fig:fig8}(a) shows the resonant wavelengths of the $p$-mode (circles) and $d$-mode (crosses). In Fig.~\ref{fig:fig8}, for the $d$-mode where the peak could not be obtained from the cross section at $k_x=0$, the mode frequency determined here was used. The plotted data in the crystalline state have been corrected by subtracting a background blueshift of 2.3\,nm. This correction clarifies that the phase transition of Sb\textsubscript{2}Se\textsubscript{3} selectively induces a redshift in the $p$-mode only.
The error bars in Fig.~\ref{fig:fig8}(a) arise from two independent sources of uncertainty: Lorentzian fitting and parabolic fitting. For the $p$-mode, only the Lorentzian fitting error is taken into account, whereas for the $d$-mode, both contributions are combined. In the Lorentzian fitting of the reflection spectrum at each $k_x$ point, the standard error of the resonant wavelength, $\sigma_{\lambda,i}$, is obtained from the diagonal elements of the covariance matrix $P_{\mathrm{cov}}$:
\begin{align}
    \sigma_{\lambda,i} = \sqrt{\mathrm{diag}(P_{cov})_0}
\end{align}
The representative Lorentzian fitting error $\bar{\sigma}_{\mathrm{Lorentz}}$ is calculated as the arithmetic mean over $N$ valid data points:
\begin{align}
    \bar{\sigma}_{\mathrm{Lorentz}} = \frac{1}{N} \sum_{i=1}^{N} \sigma_{\lambda,i}
\end{align}
For the parabolic fitting [Eq.~(\ref{def_parabolic_fit})], the residual variance $\sigma_{\mathrm{res}}^2$ is computed from the residual sum of squares $S_{\mathrm{res}}$ divided by the degrees of freedom ($N-2$):
\begin{align}
    \sigma_{\mathrm{res}} = \sqrt{\frac{S_{\mathrm{res}}}{N-2}}, \quad S_{\mathrm{res}} = \sum_{i=1}^{N}(y_i - \hat{y}_i)^2
\end{align}
The standard error of the vertex wavelength $\lambda_0$ is then given by:
\begin{align}
    \sigma_{\mathrm{vertex,quad}} = \sigma_{\mathrm{res}} \sqrt{[(X^T X)^{-1}]_{1,1}}
\end{align}
where $X$ is the design matrix for the parabolic model. The total error for the $d$-mode is obtained by combining these two independent errors via root-sum-square:
\begin{align}
    \sigma_{\mathrm{total}} = \sqrt{(\sigma_{\mathrm{vertex,quad}})^2 + (\bar{\sigma}_{\mathrm{Lorentz}})^2}
\end{align}
In the amorphous state, the spectral signal is relatively weak, resulting in error bars of 2--3\,nm. Nevertheless, even accounting for these uncertainties, the band inversion evidenced in Fig.~\ref{fig:fig8}(a) remains unambiguous. Near the band-inversion point ($\Delta R \approx 0.00675a$), the $p$- and $d$-mode dispersions exhibit an anticrossing behavior, indicating that the bandgap does not fully close. This residual gap is attributed to the coupling between the orthogonal $p$- and $d$-modes, likely induced by slight in-plane symmetry breaking arising from fabrication imperfections in the Sb\textsubscript{2}Se\textsubscript{3} nanopatterns (e.g., minor misalignment or shape asymmetry). Notably, the clear observation of this anticrossing, which would otherwise be obscured by spectral broadening in high-loss material systems, underscores the ultra-low-loss nature of Sb\textsubscript{2}Se\textsubscript{3}. Even within this anticrossing region, the measured reflection spectra [Figs.~\ref{fig:fig6}(c),(d)] largely retain the distinct radiative characteristics of the $p$- and $d$-modes.

The wavelength shift of the $p$-mode induced by the phase transition is plotted in Fig.~\ref{fig:fig8}(b). The shift is approximately 3\,nm, which is smaller than that simulated (5\,nm) in Fig.~\ref{fig:fig3}(c). This discrepancy is likely due to the small volume of Sb\textsubscript{2}Se\textsubscript{3} nanopatterns, which may affect the refractive index.
For $\Delta R = 0.00675a$, the $p$-mode exhibits a wavelength shift of approximately 10\,nm, which is significantly larger than the shift expected solely from the refractive-index change of Sb\textsubscript{2}Se\textsubscript{3} observed in other samples. This indicates that the wavelength shift is enhanced by coupling between the $p$- and $d$-modes. As shown in Fig.~\ref{fig:fig8}(a), even when the $p$- and $d$-mode wavelengths come closest, an anticrossing occurs and maintains a gap of about 7--8\,nm while preserving their distinct radiative characteristics. Under these conditions, when a small refractive-index change induces a band inversion between the $p$- and $d$-modes, the mode frequencies shift by the size of this gap in addition to their intrinsic shift.

Figure~\ref{fig:fig8}(c) plots the out-of-plane radiation Q-factor $Q_u$ for the $p$-mode. 
For $\Delta R < 0.01a$, a clear trend is observed where the out-of-plane radiation loss increases (i.e., $Q_u$ decreases) upon crystallization. This behavior arises because the increased refractive index of crystalline Sb\textsubscript{2}Se\textsubscript{3} enhances the vertical asymmetry of the slab, thereby promoting radiation into free space.
In contrast, for $\Delta R > 0.01a$, the radiation loss is dominated by the intrinsic geometric shift of the triangular holes, making the Q-factor less sensitive to the phase of Sb\textsubscript{2}Se\textsubscript{3}. These experimental trends are consistent with the numerical simulations shown in Fig.~\ref{fig:fig3}(e).
Here, although the result for $\Delta R = 0.005a$ shown in Fig.~\ref{fig:fig8}(c) face difficulties in precise fitting due to the extremely low radiation loss in the amorphous state, this does not affect the overall trend.

The total Q-factor ($Q_\mathrm{total}$) is shown in Fig.~\ref{fig:fig8}(d). In stark contrast to the GST-loaded PTI, which suffers a drop in Q-factor upon crystallization, the Sb\textsubscript{2}Se\textsubscript{3}-loaded PTI maintains a high Q-factor on the order of $10^3$ in both amorphous and crystalline states. In the small-bandgap region, the out-of-plane radiation loss is reduced, leading to a significantly weaker measurement signal as seen in Figs.~\ref{fig:fig6}(c),(d). This results in lower precision of the linewidth fitting and thus greater variation in the plotted Q-factor. However, even accounting for this variability, the difference in performance compared to highly absorptive crystalline GST is stark and unambiguous. This result definitively demonstrates that Sb\textsubscript{2}Se\textsubscript{3} overcomes the critical issue of material absorption, representing a significant advancement toward practical, low-loss, reconfigurable topological photonic devices.

\section{Conclusion}

We have successfully demonstrated an absorption-free photonic topological phase transition by integrating Sb\textsubscript{2}Se\textsubscript{3} with a silicon photonic crystal slab. Throughout the switching process, high $Q$-factors on the order of $10^3$ were maintained---a nearly order-of-magnitude improvement over GST-based devices. This demonstration was enabled by the successful realization of precise submicron-scale selective patterning of Sb\textsubscript{2}Se\textsubscript{3} on the photonic crystal. These results validate Sb\textsubscript{2}Se\textsubscript{3} as an effective solution to the material-absorption bottleneck in reconfigurable topological photonics.

Looking forward, a next step is establishing reversible operation between amorphous and crystalline states. For the reverse transition (crystalline to amorphous), laser scanning techniques offer a promising pathway. Recent demonstrations of free-form rewritable photonic circuits have shown that focused laser scanning can selectively amorphize Sb\textsubscript{2}Se\textsubscript{3} through localized melting and rapid quenching \cite{doi:10.1126/sciadv.adk1361}. This laser-based approach is directly applicable to our patterned structures: by rastering a focused laser beam across the Sb\textsubscript{2}Se\textsubscript{3} regions, we can achieve spatially selective amorphization. The small volume of our patterned Sb\textsubscript{2}Se\textsubscript{3} regions ($\sim$50\,nm thickness) should facilitate efficient heat dissipation and rapid cooling rates necessary for complete amorphization. Implementing such reversible switching would enable fully reconfigurable topological photonic devices where the topological phase can be repeatedly switched on demand. 

Beyond reversibility, the demonstrated technique of selectively integrating ultra-low-loss PCMs onto nanophotonic structures provides a versatile platform for nonvolatile control of optical properties. In contrast to GST---which intrinsically links refractive-index modulation to large loss changes---Sb\textsubscript{2}Se\textsubscript{3} enables pure index tuning due to its wide bandgap. This property is essential both for maintaining high-$Q$ factors and for preserving the effectively Hermitian character of the system, a prerequisite for defining topological invariants and ensuring robust edge states in practical implementations such as topological lasers and quantum photonic circuits. Furthermore, Sb\textsubscript{2}Se\textsubscript{3} exhibits significant third-order optical nonlinearity, with a nonlinear susceptibility $\chi^{(3)} \approx 2 \times 10^{-11}$\,esu\cite{Hamrouni2018}. This strong nonlinearity, combined with ultra-low linear absorption, opens additional pathways for combining nonvolatile reconfigurability with nonlinear optical functionalities such as all-optical switching and wavelength conversion within topological photonic platforms. Potential applications include reconfigurable waveguide circuits where topological edge states can be routed, tunable cavity arrays for quantum photonics, and exploration of non-Hermitian physics through controlled introduction of gain or loss. This study establishes a crucial foundation toward practical, reconfigurable topological photonic systems with performance suitable for real-world applications.

\section{Acknowledgements}

This work was supported by the Japan Society for the Promotion of Science (Grant numbers JP20H05641, JP21K14551, 24K01377, 24H02232, 24H00400) and JST Support for Pioneering Research Initiated by the Next Generation (JPMJSP2180). 
We acknowledge invaluable contributions from Dr. Toshiaki Tamamura, Toshifumi Watanabe, Osamu Moriwaki, and Junichi Asaoka for fabrication techniques. We also thank Shinichi Fujiura for assistance with AFM measurements.

\section{Supporting information}

Data underlying the results presented in this paper are not publicly available at this time but may be obtained from the authors upon reasonable request.

\bibliography{references}

\end{document}